\documentclass[twocolumn,english,final,showpacs,superscriptaddress,lengthcheck,prl]{revtex4}
\usepackage[T1]{fontenc}
\usepackage[latin9]{inputenc}
\usepackage{textcomp}
\usepackage{amsmath}
\usepackage{graphicx}
\usepackage{amssymb}
\def\be{\begin{equation}}
\def\ee{\end{equation}}
\def\e#1{\label{#1}\end{equation}}
\def\bea{\begin{eqnarray}}
\def\eea{\end{eqnarray}}
\def\ea#1{\label{#1}\end{eqnarray}}

\def\bem#1{\begin{mathletters}\label{#1}}
\def\eml{\end{mathletters}}

\def\4#1{{\boldsymbol{#1}}}
\def\8#1{{\widetilde{#1}}}
\def\bse{\begin{subequations}}
\def\ese{\end{subequations}}
\def\nn{\nonumber}
\makeatletter
\@ifundefined{textcolor}{}
{%
 \definecolor{BLACK}{gray}{0}
 \definecolor{WHITE}{gray}{1}
 \definecolor{RED}{rgb}{1,0,0}
 \definecolor{GREEN}{rgb}{0,1,0}
 \definecolor{BLUE}{rgb}{0,0,1}
 \definecolor{CYAN}{cmyk}{1,0,0,0}
 \definecolor{MAGENTA}{cmyk}{0,1,0,0}
 \definecolor{YELLOW}{cmyk}{0,0,1,0}
 }

\@ifundefined{definecolor}
 {\@ifundefined{definecolor}
 {\@ifundefined{definecolor}
 {\usepackage{color}}{}
}{}
}{}
\usepackage{subfigure}  
\usepackage{hyperref}   

\makeatother

\usepackage{babel}

\makeatother

\usepackage{babel}

\makeatother

\usepackage{babel}

\begin{document}

\title{Generation of macroscopic quantum-superposition states by linear coupling to a bath}

\author{D. D. Bhaktavatsala Rao}

\email{dasari@weizmann.ac.il}

\affiliation{Weizmann Institute of Science, Rehovot,
76100, Israel.}

\author{Nir Bar-Gill}

\affiliation{Department of Physics, Harvard Univeristy, Cambridge MA, USA}

\affiliation{Weizmann Institute of Science, Rehovot,
76100, Israel.}

\author{Gershon Kurizki}

\affiliation{Weizmann Institute of Science, Rehovot,
76100, Israel.}
\begin{abstract}
We demonstrate through an exactly solvable model that collective coupling to any thermal bath induces effectively nonlinear couplings in a quantum many-body (multi-spin) system. The resulting evolution can drive an uncorrelated large-spin system with high probability into a macroscopic quantum-superposition state. We discuss possible experimental realizations.
\end{abstract}

\date{\today}
\maketitle
{\it Introduction:} 
The strive for ultrafast and ultrapowerful data processing by quantum information techniques commonly relies on quantum entanglement (QE) that is induced by (direct or indirect) interactions among the particles in the system \cite{nielsen,stoler}. Yet this QE is extremely vulnerable to the environment: its fragility exponentially mounts with the strength of the system-environment (bath) coupling and the number of particles \cite{breu,davido,wine}. Alternatively, QE is realizable by collective (multipartite) dissipation, i.e., real-quanta exchange via the bath within the rotating wave approximation \cite{plenio}. In both forms of the QE, the bath effects are commonly treated within the Markov approximations \cite{breu,plenio,zoll}. Here we take an alternative approach and show through an exactly solvable model that QE can arise spontaneously from dispersive, nondissipative interactions (virtual-quanta exchange) among particles via the bath, a process unexpectedly revealed by going beyond the foregoing standard approximations.

Such bath-mediated dispersive interactions result in effectively nonlinear couplings, although we assume the system-bath coupling to be linear, as usual \cite{breu}. They are shown to grow up to an asymptotic value that is reached beyond the non-Markov (memory) time of the bath. The resulting unitary evolution can allow with high probability for the formation of entangled, macroscopic quantum superposition (MQS) states in atom or spin ensembles collectively coupled to any bosonic bath, at zero (vacuum) or nonzero temperature. The number of particles in the MQS depends on the spectral response of the bath and its temperature. This result generalizes the notion of QE via dispersive single-mode interactions in ion-traps \cite{molm} to arbitrary bath spectra and diverse scenarios.

Collective dynamics of ensembles of atoms and spins are among the few well-studied manifestations of quantum behavior on macroscopic scales.
This comes about because their quantized collective dynamics can be mapped onto that of an object in an eigenstate of angular momentum
(spin) $\vec{L}$ with large eigenvalues. Behavior of this kind is exhibited by spin-polarized ensembles in solids \cite{molmer} or by atomic ensembles
with large pseudospin that collectively emit and absorb photons \cite{dicke,scully}. Alternatively,
large spin characterizes macroscopic atomic ensembles that are entangled via interaction with a common light source \cite{polzik}.

{\it Model and dynamics:} We consider an ensemble of $N$ non-interacting spins or atomic two-level systems (TLS) that are {\it identically}, linearly coupled to a bosonic (oscillator) bath. The model is described in the collective basis by  the many-body Hamiltonian
\bea
&H=H_S+H_B+H_I,& \nn \\
&H_S=\omega_x L_x & \nn \\
&H_B=  \sum_k \omega_k b^\dagger_k b_k,~~H_I= {L}_z\sum_k \eta_k(b_k+b^\dagger_k).&
\label{eq1}
\eea
Here $b^\dagger_k, b_k$ are the creation and annihilation bosonic operators of the $k$-th bath mode, and $\eta_k$ the corresponding ($k-mode$) coupling rates.
The collective spin operators in $H_S$ and $H_I$ are $L_i = \sum_k\sigma^i_k (i=x,y,z)$, the components of the total spin $\vec{L}$ of the ensemble, $\sigma^i_k$ being the TLS operators (Pauli matrices) for the $k$-th spin-$1/2$ particle. The rotating wave approximation has not been made here. Since $H$ commutes with $L^2 = \sum_iL^2_i$, the bath interacts separately with each subspace of the system labelled by the total-spin value $l$. The $l$ values range from $0$ to $\frac{N}{2}$, if $N$ is even, and $\frac{1}{2}$ to $\frac{N}{2}$, if $N$ is odd. Hence, it is sufficient to study the interaction of the bath with a multi-level, ($2l+1$)-dimensional system. 

In general, the dynamics generated by Eq. (\ref{eq1}) is insolvable, because of the non-commutativity of $L_x$ and $L_z$. In order to circumvent this difficulty we prepare the system in an eigenstate of $L_x = \sum_k \sigma^x_k$ (a superposition of $L_z$ eigenstates) and then switch off $\omega_x$. 
This removes $H_S$ in Eq. (\ref{eq1}). The Hamiltonian dynamics then becomes exactly solvable for any bosonic bath. For a given spin-sector $l$ of the ensemble, one can then write a closed-form equation for the time-evolution operator of the system and the bath (see \cite{epaps} for more details), given by
\bea
\label{uexp}
U_l(t) =\exp\left[{-itf(t)L^2_z} + L_z\sum_k\left(\alpha_k(t)b^\dagger_k-\alpha^*_k(t)b_k\right)\right]
\eea
where the coupling spectrum of the bath (see below) determines the functions
\be
\label{fexp}
\hspace{-0.5cm}
f(t)=\frac{1}{t}\sum_k \eta^2_k(\omega_k t-\sin \omega_kt)/\omega^2_k, 
~\alpha_k(t) =  \eta_k\frac{1-{\rm e}^{i\omega_kt}}{\omega_k}.
\ee
We thus obtain a striking {\it exact} result: the bath-induced evolution is driven by both $L_z$ (linear) and $L^2_z$ ({\it nonlinear}) terms. The linear terms cause decoherence as expected. The new term $f(t)L^2_z$ in Eq. (\ref{uexp}) occurs for multipartite systems where $L^2_z = \mathcal{I} + \sum_{i\ne j}\sigma^z_i\sigma^z_j$: it is absent in the single particle case where $L^2_z = \sigma^2_z \equiv \mathcal{I}$. As discussed below, it may be interpreted as a collective analog of the Lamb shift \cite{lamb} or frequency pulling of each spin by all others via the bath (virtual quanta exchange). 

We focus on a class of initial states of the spin ensemble that can be expressed as a weighted sum of density matrices with values of $l$, $\rho(0) = \sum_l\lambda_l \rho_l(0)$. Under the above dynamics each component $\rho_l$ evolves separately. The state of the system (initially uncorrelated with the bath), at any later time, is found upon tracing over the bath degrees of freedom:
\be
\hspace{-1.2cm}
\label{rh}
\rho(t)=\sum_l\lambda_l{\rm e}^{-itf(t)L^2_z}\left\lbrace\sum_{m,m^\prime}\rho^{mm^\prime}_l(0){\rm e}^{-t\Gamma(t)(m-m^\prime)^2}|m\rangle\langle m^\prime|\right\rbrace {\rm e}^{it{f}(t) L^2_z},
\ee
where $\Gamma(t) =\frac{1}{t}\sum_k \eta^2_k(1-\cos \omega_kt)/\omega^2_k$ is the decoherence rate (discussed below) and $m$, $m^\prime$, ranging from $+l$ to $-l$, label the $L_z$ states \cite{dicke}. 

\begin{figure}[htb]
\includegraphics[width=6cm]{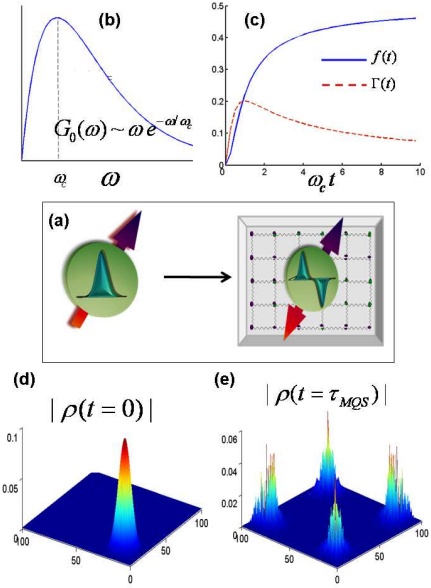}
\protect\caption
{(Color online) {\bf Bath-induced formation of a macroscopic quantum-superposition (MQS) state}. (a) A collective (large) spin naturally evolves in the bath into a MQS state. The system is a spin ensemble composed of $N=50$ particles coupled to an Ohmic bath. The cut-off frequency $\omega_c$ is chosen such that the collective coupling of the spin ensemble to the bath, $\eta = \sqrt{\sum_k \eta^2_k} \sim 0.005\omega_c$. (b) The coupling spectrum of an Ohmic bath. (c) The time-dependent functions responsible for the nonlinear Lamb-shift ($f(t)$) and decoherence ($\Gamma(t)$) dynamics of the system. (d)-(e) The absolute value of the density matrix elements, $|\rho_{mm^\prime}(t)|$ in the $L_x$ basis (Eq. (\ref{rh})) at various times. The initial state of the ensemble is a spin coherent state, $|\theta=\pi/4,\phi=0\rangle$, in  $L_x$ basis.
The presence of the $|\pm N/2\rangle\langle\mp N/2|$ off-diagonal elements (left and right most peaks) signifies the formation of such a state with high fidelity.}
\end{figure}

In keeping with the procedure leading to Eq. (\ref{uexp}), at time $t=0$ each spin is prepared in a superposition of its energy ($\sigma^z_k$) eigen states  
The total system is then initially in a product of such superposition states which is uncorrelated (unentangled) among the individual spins.
The initial uncorrelated state of the spin ensemble (all spins identical), can then be written as 
$\rho(0)=|\psi(0)\rangle\langle\psi(0)|$, where
\be
\label{psi}
\hspace{-1cm}
|\psi(0)\rangle \equiv|\theta, \phi\rangle = |\psi_1\rangle\otimes|\psi_2\rangle\cdots|\psi_N\rangle, ~|\psi_k\rangle = \cos\frac{\theta}{2}|\uparrow\rangle + \sin\frac{\theta}{2} {\rm e}^{i\phi}|\downarrow\rangle.
\ee
$\theta$ and $\phi$ are the usual Bloch-sphere angles.This state is an eigenstate of the collective spin operator $\vec{L}\cdot\hat{n}$ and $\hat{n}$ is the unit vector corresponding to the angles $\theta$ and $\phi$. 

The off-diagonal terms in Eq. (\ref{rh}) decay exponentially at the rate $\Gamma(t)(m-m^\prime)^2$, rendering multipartite coherence (entanglement) vulnerable to decoherence.
Let us, however, first consider that $\Gamma(t)$ is negligible. Then, under the nonlinear term $f(t)L^2_z$ in Eq. (\ref{uexp}), the initial uncorrelated state (\ref{psi}) evolves at prescribed times (see below) into an entangled macroscopic quantum superposition (MQS), state (analogoulsy to evolution under the nonlinear Kerr Hamiltonian \cite{stoler}): 
\be 
\label{mqs}
|\psi\rangle_{MQS} = \frac{1}{\sqrt{2}}[{\rm e}^{-i \pi/4}|\theta,\phi\rangle + {\rm e}^{i \pi/4}|\theta-\pi,\phi\rangle].
\ee

If, for example, $|\theta,\phi\rangle = |\frac{\pi}{2},0 \rangle$, then Eq. (\ref{mqs}) corresponds to a superposition of the state with $l=N/2, m=N/2$ in the $L_x$ basis, where all the spins are oriented along $+\hat{x}$ direction, and the state with $l=N/2, m=-N/2$ in $L_x$ basis, where all the spins are oriented along $-\hat{x}$ direction. The interaction with the bath in Eq. (\ref{rh}) thus transforms the initially uncorrelated state, $|+x\rangle = |\frac{\pi}{2},0\rangle$, under negligible decoherence ($\Gamma(t)\approx 0$), into a MQS which is simultaneously oriented along the $+\hat{x}$ and $-\hat{x}$ directions with $\pi/2$ relative phase 
\be
|+x\rangle\xrightarrow{H_I}\frac{1}{\sqrt{2}}[|+x\rangle - i|-x\rangle].
\ee
This is a macroscopic $GHZ$-like state in which all the $N$ spins are maximally entangled. As is known, the parity of $N$ plays a role \cite{stoler,plenio}: only even $N$ strictly result in GHZ states, although for large $N$ the deviation from perfect GHZ states for odd $N$ is negligible.

The MQS will form according to Eq. (\ref{fexp}), (\ref{rh}) at times when $tf(t)=(2n+1)\pi/2$ ($n$ being an integer) \cite{stoler}. The earliest formation time of the MQS is then
\be
\label{tauexp}
\tau_{MQS} \equiv t =  \frac{\pi}{2f(t)}.
\ee
The time at which such a state forms is independent of $N$. Other entangled states (``Schroedinger kittens'') will form at intermediate times \cite{stoler,aver}. 

Since the bath spectrum is continuous, the decoherence $\Gamma(t)\ne 0$ is unavoidable and the formation of perfect MQS states is generally not possible. Yet, the formation of high-purity states i.e., the accomplishment of Eq. (\ref{mqs}) at $\tau_{MQS}$ (Eq. (\ref{tauexp})) with high probability may still be possible, as argued below.
The condition for the quantum-superposition to survive decoherence is, from Eqs. (\ref{rh})-(\ref{tauexp}), that the MQS forms faster than it decays:
\be
\label{tau}
 \tau_{MQS}\bar{\Gamma} N^2 < 1.
\ee
where $\bar{\Gamma} N^2$ is the upper limit of the time-averaged decay rate $\Gamma(t)(m-m^\prime)^2$ of the off-diagonal (coherence) elements in Eq. (\ref{rh}).

{\it Dependence of MQS formation on bath spectrum and temperature}:
What determines whether the condition Eq. (\ref{tau}) for MQS formation can be satisfied ? This requires that $f(t)$ strongly exceeds $\Gamma(t)$ at the time $\tau_{MQS}$. 
The functions $f(t)$ and $\Gamma(t)$ in Eqs. (\ref{fexp}), (\ref{rh}), (\ref{tauexp}) are then expressed as
\bea
\label{fg}
f(t) &=& \frac{1}{t}\int^\infty_0 G_0(\omega) \frac{\omega t-\sin \omega t}{\omega^2},\nn \\
\Gamma(t) &=& \frac{1}{t}\int^\infty_0 G_T(\omega) \frac{1-\cos \omega t}{\omega^2},
\eea
where $G_0(\omega)$ and $G_T(\omega)$ are, respectively the zero-temperature and finite-temperature coupling spectra of the bath. The temperature-dependent coupling spectrum of the bath is defined as \cite{breu, kurizki} $G_T(\omega)= \sum_k \eta^2_k \int d\omega \delta(\omega-\omega_k){\rm coth}\beta\omega$, where $\beta$ is the inverse temperature of the bath. 
It is seen from Eq. (\ref{fg}) that $f(t)$ is related to the bath-induced Lamb shift \cite{lamb} (the real part of the zero-temperature bath susceptibility) and $\Gamma(t)$ to the bath-induced decoherence rate (the imaginary part of the temperature-dependent susceptibility \cite{breu,lamb}). 
Hence, the feasibility of condition (\ref{tau}), which requires $f(\tau_{MQS}) \gg \Gamma(\tau_{MQS})$, is determined by the bath coupling spectrum and temperature.

It is advantageous for the satisfaction of condition (\ref{tau}) that the $\tau_{MQS}$ exceeds the non-Markov time scale for the following reason. The decoherence rate $\Gamma(t)$  is  time-dependent in the non-Markov regime of $t < t_c$, where $t_c$, the correlation (memory) time of the bath is the inverse width of $G_T(\omega)$ \cite{kurizki}.
At sufficiently low temperatures, $\Gamma(t)$ is drastically reduced in the Markovian limit ($t \gg t_c$) as opposed to its fast initial non-Markovian increase: $\Gamma(t\ll t_c) \gg \Gamma(t\rightarrow\infty) = \Gamma_M$ (Fig. 1(c), 2(a)). This comes about since $\Gamma(t)$ that initially has contributions from all the bath modes, $\int G_T(\omega)d\omega$, subsequently decreases, as the bath mode-oscillators go out of phase in the Markov regime \cite{kurizki}. 
By contrast, $f(t)$ increases in the course of the transition from the non-Markov to the Markov regime, where it settles at its long-time value $|f(t\rightarrow \infty)| = |f_M| \gg |f(t\ll t_c)|$. Since the MQS-state formation time $\tau_{MQS}$ is typically longer than the bath correlation time $t_c$, MQS encounters a much lower $\Gamma \approx \Gamma_M$, and much higher $f \approx f_M$, than those encountered in the non-Markov regime, thereby helping satisfy (\ref{tau}).

At sufficiently high temperatures we attain the regime $\Gamma_M \ge f_M$ where the formation of MQS is not possible. For an Ohmic bath the condition $f_M > \Gamma_M$ is satisfied when the cutoff frequency (energy) $\omega_c$ is larger than the thermal energy, $\omega_c > k_BT$ $(\hbar=1)$ \cite{epaps}.

These trends can be seen from Fig. 1(c) for an Ohmic bath with coupling spectrum $G_0  = \alpha \omega {\rm e}^{-\omega/\omega_c}$ and Fig. 2(a), for a Lorentzian bath with coupling spectrum $G_0(\omega) = \alpha \frac{\omega^2_c}{\omega^2_c + (\omega-\omega_0)^2}$.
Different bath spectra having the same width, i.e.,the same inverse correlation time $1/t_c$, may still have different $f_M$ values and different $\Gamma_M$ in the long-time Markov limit and hence yield MQS with different purities at $\tau_{MQS}$ according to (Eq. (\ref{tau})). 

{\it Spin ensemble coupled to a phonon bath: }
As an experimentally feasible scenario, consider a localized aggregate of $N$ weakly interacting spin-$1/2$ particles (e.g. electron spin ensembles in fullerenes \cite{molmer, fuller} or quantum dots \cite{imam}) that undergo dephasing via identical coupling to phonons in a lattice, of longer wavelength than the aggregate size . Hence, they conform to the model of Eq. (\ref{eq1}).
For an Ohmic phonon spectrum with the Debye cutoff $\omega_D$ we find \cite{epaps} $f_M = \alpha \omega_D$, and $\Gamma_M = \alpha/\beta$, where $\beta$ is the inverse temperature and $\alpha < 1$ is a dimensionless coupling constant. Condition (\ref{tau}) may be satisfied at temperatures below $1mK$, for $f_M \sim \omega_D \sim 10^{13} Hz \gg \Gamma_M \ge MHz$ to obtain a quantum-superposition (GHZ state) with $N \sim 300$ spins with high fidelity (Fig. 1) at $\tau_{MQS}$.

{\it Atomic ensemble coupled to a cavity}:
Another example we consider is a non-interacting atomic gas coupled to a single-mode cavity \cite{rempe} (photonic bath). Whereas for isotropic spin ensembles the quantization axes in (1) are arbitrary, this is not the case for two-level atoms (TLS) i.e., pseudospins \cite{dicke,polzik,scully}. There the energy  splitting
of $H_S$ levels is represented by $H_S = \omega_z L_z$. Correspondingly, $L_z-$coupling of atoms to a photonic bath causes pure cooperative dephasing without population exchange, while $L_x$ coupling  to the bath causes {\it cooperative population exchange} or relaxation. In the latter case, we have to initialize the system in a $L_z$-eigenstate and then bring each TLS to degeneracy i.e. set $\omega_z = 0$ by applying Zeeman shifts.
Once the (two-level) atoms are prepared in degenerate Zeeman states ($H_S = 0$) one can induce $L_x$ coupling between the cavity and the atomic ensemble by a two-photon Raman process . The collective nonlinear evolution can thus generate MQS of the atomic ensemble. For finite cavity linewidth, the cavity acts as a Lorentzian bath, which results in  dynamics similar to Fig. 2(a). The decoherence rate $\Gamma_M$ is determined by the zero-frequency coupling strength to the cavity $G_T(0)$, or may be induced by an external low-frequency noise. This may limit the size of MQS allowed by Eq. (\ref{tau}) to $N \sim 100$, for an atom-cavity coupling strength $\eta \sim 1MHz$.

\begin{figure}[htb]
\includegraphics[width=8cm]{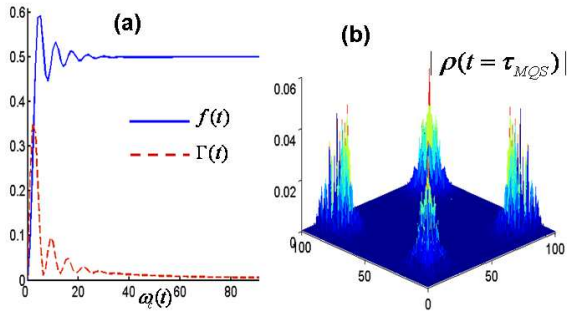}
\protect\caption
{(Color online) {\bf Interaction with a cavity mode}. As in Fig. 1, for a cavity with Lorentzian coupling spectrum with width $\omega_c$ centered at $10\omega_c$. (a) The $f(t)$ and $\Gamma(t)$ functions. (b) The formation of macroscopic quantum-superposition for cavity with Lorentzian lineshape at $\tau_{MQS} = 10^2/\omega_c$.}
\end{figure}
{\it Conclusions}: We have used an exactly solvable model to reveal the unexpected entangling dynamics of a system with large angular momentum that is linearly, collectively coupled to a thermal bosonic bath. The intriguing consequence is that a commonly occurring finite-temperature environment may naturally induce rather than impede the formation of macroscopic quantum-superposition (MQS) states. Such counterintuitive bath-induced effects change our fundamental perspective of non-classicality in open quantum systems by identifying a broad class of natural entanglement. On the applied side, the feasibility of high-fidelity entangled states with $N \geq 100$ may be a starting point to the advancement of 
one-way quantum computing \cite{brigel}.

Equations (\ref{uexp}), (\ref{rh}), (\ref{tauexp}), (\ref{tau})  are the main  results of this exact analysis. They imply novel dynamic features in open multispin systems at low temperatures: (i) Non-linear ($L^2_z$) terms induced by a bosonic bath can dominate the evolution of multi-spin systems. By contrast, for a single spin-$1/2$ (two-level) system, $L_{z(x)}^2 = \sigma^2_{z(x)} \equiv \mathcal{I}$, the nonlinear term only yields an overall phase and does not affect the dynamics. (ii) While the non-Markovian bath dynamics affects the squeezing of an initial coherent state, its Markovian dynamics governs the formation of a high purity MQS.
(iii) The design of a bath-coupling spectrum that determines a high ratio of the Lamb shift $f_M$ to the decoherence rate $\Gamma_M$, can play a crucial role in allowing the formation of such macroscopic quantum-superposition states.
(iv)Their purity may be further improved using quantum control techniques to dynamically modulate the system levels \cite{universal_decay} and thus modify the bath effects.
(v) Another intriguing consequence of the analysis is that only a bosonic bath induces such nonlinear dynamics, whereas a fermionic bath generates also higher-power nonlinearities which could be unfavorable for the generation of high-fidelity cat states.

The support of EC (MIDAS project), DIP, GIF and the Humboldt-Meitner Award is acknowledged by G. K. \\

\end{document}